 \documentstyle[prd,aps,eqsecnum,floats,epsfig]{revtex}
  \tighten
\begin{document}
%
%....................................................................%
%......   unCOMMENT THESE TO GET TWO COLUMNS:
 \twocolumn[
 \hsize\textwidth\columnwidth\hsize\csname@twocolumnfalse%
 \endcsname
%....................................................................%
%
%....................... Title, Authors .............................%
\title{Electromagnetic form factors of light vector mesons}
%.... PREPRINT NUMBER ...............................................%
 \vspace*{-.9cm}\hspace*{\fill}
 \parbox{8.0cm}{ % {\sf http://xxx.lanl.gov/abs/nucl-th/9806025} \\
		{\hspace*{\fill} \sf ADP-98-24/T300} \\
		{\hspace*{\fill} \sf FSU-SCRI-98-054} 
 } \vspace*{+0.9cm}
%..................................................................
%
\author{F.T.~Hawes}
\address{Department of Physics and Mathematical Physics, \\
        University of Adelaide, South Australia 5005, Australia}
\author{M.A.~Pichowsky}
\address{Department of Physics and 
	Supercomputer Computations Research Institute, \\
        Florida State University, Tallahassee, FL 32306-4130, USA
	}
\date{\today}
\maketitle
%....................................................................%
%....................................................................%
\begin{abstract}
The electromagnetic form factors
$G_E(q^2)$, $G_M(q^2)$, and $G_Q(q^2)$, 
charge radii, magnetic and quadrupole moments, and decay widths 
of the light vector mesons $\rho^+$, $K^{*+}$ and $K^{*0}$ 
are calculated 
in a Lorentz-covariant, Dyson-Schwinger equation based model using
algebraic quark propagators that incorporate confinement, 
asymptotic freedom, 
and dynamical chiral symmetry breaking,
and vector meson Bethe-Salpeter amplitudes closely related
to the pseudoscalar amplitudes obtained
from phenomenological studies of $\pi$ and $K$ mesons.
Calculated static properties of vector mesons include the charge radii and
magnetic moments: 
    $\langle r^2_{\rho+}  \rangle^{1/2} =$  0.61~fm,
    $\langle r^2_{K^{*+}} \rangle^{1/2} =$  0.54~fm, and
    $\langle r^2_{K^{*0}} \rangle       =$ -0.048~fm$^2$;
    $\mu_{\rho+}  =$  2.69,
    $\mu_{K^{*+}} =$  2.37, and
    $\mu_{K^{*0}} =$ -0.40.
The calculated static limits of the $\rho$-meson form factors are similar
to those obtained from light-front quantum mechanical calculations, 
but begin to differ above $q^2$ = 1~GeV$^2$ due to the dynamical evolution
of the quark propagators in our approach.
\end{abstract}
%.....................................................................
%......  unCOMMENT THIS TO GET TWO COLUMNS:
 ]
%..................................................................%
%..................................................................%
\newcommand{\Fig}[1]{Fig.~\protect\ref{#1}}
\newcommand{\Figs}[1]{Figs.~\protect\ref{#1}}
\newcommand{\Ref}[1]{Ref.~\protect\cite{#1}}
\newcommand{\Eq}[1]{Eq.~(\protect\ref{#1})}
\newcommand{\Eqs}[1]{Eqs.~(\protect\ref{#1})}
\renewcommand{\-}{\!-\!}
\renewcommand{\+}{\!+\!}
\newcommand{\T}{{\bf T}}
\newcommand{\sfrac}[2]{\mbox{$\textstyle \frac{#1}{#2}$}}
\newcommand{\nn}{\nonumber\ }        
%.........................................................................
%.........................................................................
\section{Introduction}

The electromagnetic (EM) form factors of hadrons provide an
important tool for the study of bound states in QCD.  
There have been numerous experimental and theoretical studies
of the $\pi$- and $K$-meson EM form factors
$F_{\pi}(q^2)$ and $F_{K}(q^2)$
which have improved our understanding of the dynamics
of quarks and gluons in hadrons \cite{pi_expt,Roberts,Burden}.
Fewer studies have been undertaken for the three independent vector-meson EM
form factors $F_1(q^2)$, $F_2(q^2)$ and $F_3(q^2)$,
even though they would provide information about bound-state
dynamics that is unobtainable from pseudoscalar-meson studies alone.  
In particular, EM gauge invariance constrains the {\em static limits} 
($q^2 = 0$) of $F_\pi(q^2)$, $F_K(q^2)$ and $F_1(q^2)$ to equal
the charge of the $\pi$, $K$ and vector mesons, respectively. 
Hence, the values of these form factors at $q^2 = 0$ provide {\em no}
dynamical information. 
The lack of such a constraint on $F_2(q^2)$ and $F_3(q^2)$
means that vector mesons have non-trivial static properties 
(their electric and magnetic moments).  These provide information about
bound-state dynamics in the limit that the vector meson does not recoil
from the EM probe.

Light vector-meson EM form factors have received less attention
than their pseudoscalar counterparts because their experimental
determination is more difficult. 
The short lifetime of vector mesons 
(exemplified by the ratio of the $K^*$- and $K$-meson lifetimes,
$\tau_{K^*} / \tau_{K} \approx 10^{-14}$)  
has so far excluded the possibility of a direct measurement.
However, indirect measurements of the static limits
of the EM form factors are possible.
For example, radiative decays of vector mesons,
such as $\rho^+ \rightarrow \pi^+ \pi^0 \gamma$,
may be used to obtain their magnetic moments \cite{LopezCastro},  
and charge radii may be inferred from the $t$-dependence
of cross sections for diffractive vector-meson electroproduction.
Such measurements are currently accessible at TJNAF, DESY, and FermiLab. 

Herein, the EM form factors for $\rho^+$, $K^{*+}$ and
$K^{*0}$ mesons are calculated using model confined-quark propagators
from \Ref{Burden} and vector-meson Bethe-Salpeter (BS) amplitudes that are
a simple extension of the pseudoscalar BS amplitudes also developed in
\Ref{Burden}. 
In the present study, the ramifications of employing vector-meson BS
amplitudes that are closely related to their pseudoscalar counterparts
are explored.
The parameters introduced for the model $\rho$- and $K^*$-meson BS
amplitudes are determined from a fit to the experimental values 
for the $\rho^+ \rightarrow \gamma \pi^+$, $\rho \rightarrow e^+ e^-$, 
and $K^{*+} \rightarrow \gamma K^+$ decay widths.
The calculated vector-meson EM form factors, associated moments and charge
radii are then compared to those obtained by other theoretical studies
\cite{Keister,deMelo}. 
The motivation of this study is to develop simple model BS amplitudes
for light vector mesons and to provide experience and insight necessary
for future studies of two- and three-body bound states.
Such model amplitudes are important for a wide range
of phenomenological applications including
studies of effects outside the standard model \cite{Hecht}, 
diffractive vector-meson electroproduction \cite{Pichowsky},
and $\rho$-$\omega$ mixing \cite{Mitchell}.
\nopagebreak
%
%.........................................................................
%.........................................................................
\section{Model Schwinger functions}
The use of model confined-quark propagators and meson BS amplitudes, 
based on studies of the Dyson-Schwinger equations of QCD, has 
proved efficacious to the study of hadron phenomena.
The large number of observables correlated within the framework using only
a few model Schwinger functions allows one to constrain the model forms of
these functions.  Such constraints make the framework highly predictive,
which is tested in applications to processes outside the domain
in which the model Schwinger functions were originally constrained.    
One example of this is found in \Ref{Pichowsky}, in which $u$-, $d$- and
$s$-quark propagators and $\pi$- and $K$-meson BS amplitudes, developed in
low-energy studies of $\pi$ and $K$ meson decays and EM form factors, were
employed in a study of high-energy ($\sqrt{s} \geq $ 10~GeV) $\pi$-nucleon
and $K$-nucleon diffractive scattering and Pomeron exchange.

In the following, the $s$-quark propgator parameters, originally obtained
in \Ref{Burden}, are recalculated by fitting to $K$-meson observables and
quark condensates.  This fitting proceedure leads to an $s$-quark
propagator and $K$-meson BS amplitude that is more
consistent with recent numerical studies of the Bethe-Salpeter equation
\cite{Maris}. 

%.........................................................................
%.........................................................................
\subsection{Quark propagators and pseudoscalar BS amplitudes}
Numerical studies \cite{Frank} of the coupled, Dyson-Schwinger and
Bethe-Salpeter equations in the $u$- and $d$-quark sectors obtain solutions
similar in form to those developed in \Ref{Burden}. Recent and
more extensive studies (which include the $s$-quark sector) \cite{Maris}
obtain $K$-meson BS amplitudes that are qualitatively different than
the model forms developed in \Ref{Burden}.  
To improve the connection between the model Schwinger functions employed
herein and the solutions obtained from numerical studies of the
Dyson-Schwinger and Bethe-Salpeter equations, the $s$-quark 
propagator parameters are refitted to $K$-meson observables. 

The new $s$-quark propagator is obtained by starting with the model
$u$-quark propagator and then allowing only 2 of the parameters to vary
from their $u$-quark values.
The parameters are determined by fitting the $K$-meson weak decay constant
$f_K$ and mass $m_K$, and by demanding that the $s$-quark in-vacuum and
$s$-quark in-kaon condensates and current $s$-quark mass $m_s$ are the
same as those obtained in \Ref{Maris}.
The first application of the new $s$-quark parametrization is in a
calculation of the charge radii of the neutral- and charged-$K$ mesons.  

The form of the model dressed-quark propagators is based on numerical
studies of the quark Dyson-Schwinger equation with a realistic model-gluon
propagator \cite{Frank}.  
In a general covariant gauge, the quark propagator is written as
$S_f(k) = - i \gamma \cdot k \sigma_V^f(k^2) + \sigma_S^f(k^2)$, 
and is well-described by the following parametrization:
\begin{eqnarray}
  \bar{\sigma}^f_S(x) & = & 
    \left(b^f_0 + b^f_2 {\cal F}[ \epsilon x]  \right)
    {\cal F}[b^f_1 x] {\cal F}[b^f_3 x]
      \nonumber\\    
   & &  
    + 2 \bar{m}_f {\cal F}\big[ 2(x+\bar{m}_f^2) \big] 
	+ C_{{m}_f} e^{-2x}\:, \nonumber\\
  \bar{\sigma}^f_V(x) & = & 
  \frac{ 2(x+\bar{m}_f^2) - 1 + e^{-2(x+\bar{m}_f^2)} }
       { 2(x+\bar{m}_f^2)^2 }\:, \label{QuarkProp}
\end{eqnarray}
where
\begin{equation}
{\cal F}[x] = \frac{ 1 - e^{-x} }{x},
\end{equation}
with $x = k^2/\lambda^2$, $\bar{\sigma}^f_{S} = \lambda \sigma^f_S$,
$\bar{\sigma}^f_V = \lambda^2 \sigma^f_V$, 
$\bar{m}_f = m_f / \lambda$,
$\lambda =$ 0.566~GeV, and $\epsilon = 10^{-4}$.
This dressed-quark propagator has {\em no} Lehmann representation and hence
describes the propagation of a confined quark.
It also reduces to a bare-fermion propagator when its momentum is
large and spacelike, in accordance with the asymptotic behavior expected
from perturbative QCD (up to logarithmic corrections).

The parameters for the $u$- and $d$-quark propagators are assumed to be
the same ($SU(2)$-flavor symmetric).  Their values were determined in
\Ref{Burden} by performing a $\chi^2$ fit to a range of $\pi$- and
$K$-meson observables. 
The resulting parameters are given in Table~\ref{Tab:QuarkParam}.

%.........................................................................
%..................................................................
\begin{table}
\begin{center}
\begin{tabular}{l|lllllc}
% \hline  
$f$/meson& $b^f_0$ & $b^f_1$ & $b^f_2 $ & $b^f_3$ & $C^f$ 
				&$m_f$ [MeV] \\  \hline
$u,d$ &   0.131 &   2.900 &   0.603  & 0.185 &  $\;\;$0  &   5.1 \\
$\pi$ &   same  &   same  &   same   & same  &  0.121    & $0\;\;\;$ \\
$\rho$&   0.044 &   0.580 &   same   & 0.462 &  $\;\;$0  & $0\;\;\;$ \\ \hline
$ s $ &   0.105 &   2.900 &   0.540  & 0.185 &  $\;\;$0  & 130.0 \\
$ K $ &   0.322 &   same  &   same   & same  &  $\;\;$0  & $0\;\;\;$ \\
$K^*$ &   0.107 &   0.870 &   same   & 0.092 &  $\;\;$0  & $0\;\;\;$ \\
\end{tabular}
\caption{Confined-quark propagator parameters for $u$, $d$, and $s$
quarks from \Ref{Burden}. 
An entry of ``same'' for a meson indicates that the value of the 
parameter is the same as for the quark flavor with which it is grouped.}
\label{Tab:QuarkParam}
\end{center}
\end{table}
%..................................................................
%.........................................................................

In the chiral limit ($m_f = 0$), it follows from Goldstone's theorem that
the pseudoscalar-meson BS amplitude $\Gamma_{\cal P}(k ; p)$ is completely
determined by the dressed quark propagator $S_f(k)$ \cite{Maris}.
If the amplitude is taken to be proportional to lowest Dirac moment,
$\gamma_5$, it is given by 
\begin{equation}
  \Gamma_{\cal P}(k ; p)
  = \gamma_5 \:
    \frac{ B_{\cal P}(k^2) }{f_{\cal P}} 
  \label{PiBSAmp}
\end{equation}
for ${\cal P} =$ $\pi$- or $K$-mesons, where $B_{\cal P}(k^2)$ is the
Lorentz-scalar function appearing in the quark inverse propagator
$S^{-1}_{f}(k) = i \gamma \!\cdot\! k \, A_f(k^2) + B_f(k^2)$.
Here, $f_{\cal P}$ is the pseudoscalar-meson weak-decay constant and can
be evaluated from the BS normalization\footnote 
{
  A formulation of quantum field theory is employed
  using the Euclidean metric $\delta_{\mu \nu} = {\rm diag}(1,1,1,1)$ with 
  the convention that {\em spacelike} momentum $q_{\mu}$ satisfies
  $q^{2} = q \cdot q = q_{\mu} q_{\mu} > 0 $, and 
  Dirac matrices $\gamma_{\mu}$ that satisfy:
  $\gamma_{\mu}^{\dag} = \gamma_{\mu}$ and
  $\{ \gamma_{\mu},\gamma_{\nu} \} = 2 \delta_{\mu \nu}$.
}:
\begin{eqnarray}
f^2_{\cal P} &=& \frac{p_{\mu}}{p^2} 
N_c {\rm tr} \int \!\frac{{\rm d}^4 k}{(2\pi)^4} 
B^2_{\cal P}(k^2) 
\nn \\
& & \times
\bigg( \frac{\rm d\;}{{\rm d}p_{\mu}} S_s(k\+\sfrac{1}{2}p) \gamma_5
S_u(k\-\sfrac{1}{2}p) \gamma_5
\nn\ \\
& &+ 
S_s(k\+\sfrac{1}{2}p) \gamma_5
\frac{\rm d\;}{{\rm d}p_{\mu}} S_u(k\-\sfrac{1}{2}p)
\gamma_5 \bigg)
\label{fK}
,
\end{eqnarray}
where $S_f(k)$ is the propagator for a quark with flavor $f$,
$B_{\cal P}(k^2)$ is from the BS amplitude for the pseudoscalar, 
${\cal P}= \pi$ or $K$ meson in \Eq{PiBSAmp}, 
with total momentum $p$ and relative
quark-antiquark momentum $k$, $N_c = 3$, and ${\rm tr}$ denotes a trace
over the Dirac indices.

Using the definition in \Ref{Roberts}, the in-vacuum quark condensate
obtained from the model quark propagator of \Eq{QuarkProp} is 
\begin{equation}
\langle \bar{q}^f {q^f} \rangle^{\mu}_{\rm vac}
= - \lambda^3 \left( \ln \frac{\mu^2}{\Lambda^2_{\rm QCD}} \right)
\frac{3}{4\pi^2} \frac{b_0^f}{b_1^f b_3^f}.
\label{VacCond}
\end{equation}

The $s$-quark parameters are chosen as follows.
The magnitude of the $s$-quark condensate in the
vacuum is taken to be 80\% that of the $u$ quark.  The scales $b^f_1$ and
$b^f_3$ in \Eq{QuarkProp} are flavor independent 
($b^s_1 = b_1^u$ and $b_3^s = b_3^u$) so that 
\Eq{VacCond} fixes the value of $b_0^s = 0.80 \; b_0^u$.
A relation for the ``in-kaon'' quark condensate 
$\langle \bar{q}^s q^s \rangle^{\mu}_{K}$ 
is obtained from \Eq{VacCond} with the replacement 
$b_0^s \rightarrow b_0^K$.  
The requirement that this condensate and the current-quark mass $m_s$ have
the same values as obtained in \Ref{Maris} at the renormalization point
$\mu =$ 1~GeV, determines $b_0^K=$ 3.07 $b_0^s$.

The $K$-meson mass $m_K$ is determined from the relation:
\begin{equation}
f_K^2 m^2_K = - \langle \bar{q}^s q^s \rangle_{\mu^2}^{K} (m_u + m_s)
.
\end{equation}
The value of $b^s_2$ is obtained by performing a $\chi^2$ fit to the
experimental values of $f_K$ and $m_K$.
The resulting parameters are given in Table~\ref{Tab:QuarkParam}, the
resulting weak decay width, $K$-meson mass and $s$-quark condensates are
given in Table~\ref{Tab:BSAmpParam}.

%.........................................................................
%..................................................................
\begin{table}
\begin{center}
\begin{tabular}{c|cc}
 		&  this study  		& experiment \\ \hline
$f_K$ 	 	&  0.112  		& 0.113 $\pm$ 0.001    \\
$m_K$   	&   0.491  		& 0.494 $\pm$ 0.016    \\
$- \langle \bar{q}^s q^s \rangle_0^{{1/}{3}} $ 
		&  0.198  & 0.175 -- 0.205 \\
$- \langle \bar{q}^s q^s \rangle_{K}^{{1/}{3}} $
		& 0.289  & --- \\
$\langle r^2_{K^+} \rangle^{{1/}{2}} $
		& 0.528 		& 0.582 $\pm$ 0.041  \\
$\langle r^2_{K^0} \rangle$
		& -0.032 		& -0.054 $\pm$ 0.026  
\end{tabular}
\caption
{
Calculated $K$-meson observables and $s$-quark condensates evaluated at
$\mu =$ 1~GeV. 
All values are given in units of GeV, except
$\langle r^2_{K^+} \rangle^{{1/}{2}}$ is in fm, 
and $\langle r^2_{K^0} \rangle$ is in fm$^2$ \protect\cite{pi_expt}. 
Experimental values given for condensates are those obtained from
theoretical studies (see \Ref{Burden}).
} 
\label{Tab:BSAmpParam}
\end{center}
\end{table}
%..................................................................
%.........................................................................

%%............................................................
\begin{figure}[t]
\centering{\ 
\mbox{\ 
 \epsfig{figure=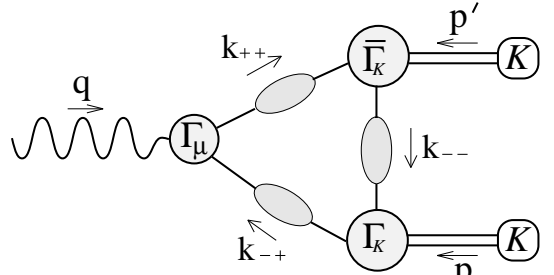,width=7.0cm}}}
\caption{
One contribution to the proper $\gamma K \bar{K}$ vertex.
}
\vspace{1.0ex}
\label{Fig:GKK}
\end{figure}
%%............................................................

The first application that will serve as a test of the new parametrization
of the $s$-quark propagator is a calculation of the charged and neutral
$K$-meson charge radii.  
The charge radius is related to the $K$-meson EM form factor by 
\begin{equation}
\langle r_{K^+}^2 \rangle  = - 6 \lim_{q^2 \rightarrow 0}
	\frac{\rm d\;}{{\rm d} q^2} F_{K^+}(q^2)
\label{Def:Radius}
,
\end{equation}
and $F_{K^{+}}(q^2)$ is obtained from the $\gamma K^+ K^-$ vertex:  
\begin{equation}
\Lambda^{K^+}_{\mu}(p,p^\prime) = 
	\sfrac{1}{2}(p\-p^{\prime})_{\mu} \; F_{K^+}(q^2)
	+\sfrac{1}{2}(p \+ p^{\prime})_{\mu} \; G_{K^+}(q^2)
,
\end{equation}
where $p$ ($p^\prime$) is the four-momentum of the initial (final) $K$
meson, $q = - p - p^{\prime}$ is the four-momentum of the photon.
In this elastic ($p^2 = p^{\prime 2}$) process, the form factors depend
only on $q^2$ and $G_{K}(q^2) = 0$ for all $q^2$. 
In the impulse approximation, the $\gamma K^+ K^-$ vertex is given
by:
\begin{equation}
\Lambda^{K^+}_{\mu }(p,p^\prime) = 
 \sfrac{2}{3} \Lambda_{\mu }^{u\bar{s}}(p,p^\prime)
+ \sfrac{1}{3} \Lambda_{\mu}^{\bar{s}u}(p,p^\prime),
	\label{LambdaKDef}
\end{equation}
with
\begin{eqnarray}
  \Lambda_{\mu}^{u\bar{s}}(p,p^\prime) &=&
  2 N_c {\rm tr}\int \frac{{\rm d}^4k}{(2\pi)^4}
  S_u(k_{++}) i \Gamma_{\mu}(k_{++},k_{-+}) 
\nn \\   & & \times 
  S_u(k_{-+}) \bar{\Gamma}_{K}(k\-\sfrac{1}{2}q;\-p^\prime) 
\nn \\   & & \times 
  S_s(k_{--}) {\Gamma}_{K}(k;p) , \label{LambdaKU}\\
   \Lambda_{\mu}^{\bar{s}u}(p,p^\prime) &=&
   2 N_c {\rm tr}\int \frac{{\rm d}^4k}{(2\pi)^4} 
   S_s(k_{++}) i \Gamma_{\mu}(k_{++},k_{-+})
\nn \\   & & \times 
   S_s(k_{-+}) {\Gamma}_{K}(k\-\sfrac{1}{2}q;-p^\prime) 
\nn \\   & & \times 
S_u(k_{--}) \bar{\Gamma}_{K}(k;p), \label{LambdaKS}
\end{eqnarray}
where $k_{ab} = k \+ \sfrac{a}{2} q \+ \sfrac{b}{2} p$,
$N_c = 3$, 
$\Gamma_{K}(k;p)$ is the pseudoscalar $K$-meson BS amplitude, 
$\Gamma_{\mu}(k,k^\prime)$ is the dressed quark-photon vertex,
$S_f(k)$ is the dressed propagator for a confined-quark of flavor $f$
and ``${\rm tr}$'' denotes a trace over Dirac indices.
The adjoint pseudoscalar-meson BS amplitude is given by
$\bar{\Gamma}_{K}(k;-p) = C^T \Gamma_{K}^T(-k;-p) C$, where 
$ C = \gamma_2 \gamma_4$ and $T$ denotes the transpose of Dirac indices.
The proper $\gamma K^{0} \bar{K}^{0} $ vertex is given by 
Eqs.~(\ref{LambdaKDef})--(\ref{LambdaKS}) with the relabeling 
$u \rightarrow d$ and the replacement of the coefficient $\sfrac{2}{3}$ by 
$-\sfrac{1}{3}$ in \Eq{LambdaKDef}.
For comparison, the proper $\gamma \pi^+ \pi^-$ vertex would be given by
Eqs.~(\ref{LambdaKDef})--(\ref{LambdaKS}) with the relabeling 
$K \rightarrow \pi$ and $s \rightarrow d$.
The diagram corresponding to \Eq{LambdaKU} or (\ref{LambdaKS}) is shown in
\Fig{Fig:GKK}.

Evaluation of the EM form factors from
Eqs.~(\ref{LambdaKDef})--(\ref{LambdaKS}) requires an explicit form for
the quark-photon vertex $\Gamma_{\mu}(k,k^\prime )$. 
The quark-photon vertex $\Gamma_{\mu}(k,k^\prime )$ transforms under charge
conjugation, space and time inversion in a manner similar to the bare
vertex $\gamma_{\mu}$ and must satisfy the Ward-Takahashi identity (WTI).
The most general 
quark-photon vertex, free of kinematic singularities, that fulfills these
requirements is written as $\Gamma^{\rm T}_{\mu}(k,k^{\prime}) 
+ \Gamma^{\rm BC}_{\mu}(k,k^{\prime})$ 
\cite{Bashir}, where $\Gamma^{\rm T}_{\mu}(k,k^{\prime})$ is
transverse to the photon momentum $q_{\mu} = (k - k^{\prime})_{\mu}$ and
$\Gamma^{\rm BC}_{\mu}(k,k^{\prime})$, determined from the WTI, is
\cite{Ball}: 
\begin{eqnarray}
  i \Gamma^{\rm BC}_{\mu}(k,k^{\prime}) &=& 
  \sfrac{i}{2} \gamma_{\mu} f_1
  + \sfrac{i}{2} (k+k^{\prime})_{\mu} 
	\gamma \cdot ({k+k^{\prime}}) f_2
\nn\ \\ 
& & + (k+k^{\prime})_{\mu} f_3 ,
\label{GammaBC}
\end{eqnarray}
with $f_1 = A_f(k^2) + A_f(k^{\prime 2})$, 
$f_2 = (A_f(k^2)-A_f(k^{\prime 2}))/(k^2-k^{\prime 2})$,
$f_3 = (B_f(k^2)-B_f(k^{\prime 2}))/(k^2-k^{\prime 2})$, 
where $A_f(k^2)$ and $B_f(k^2)$ are the Lorentz invariant functions
in the dressed-quark inverse propagator 
$S_f^{-1}(k) = i \gamma \cdot k A_f(k^2) + B_f(k^2)$.
The addition of transverse contributions to the quark-photon vertex
in $\Gamma^{\rm T}_{\mu}(k,k^{\prime})$ are unimportant for the
calculation of 
observables at spacelike momenta \cite{NoGammaT}; 
they vanish as both $k^2$ and $k^{\prime 2}$ become large and
spacelike in accordance with the asymptotic freedom of QCD 
and vanish in the limit that $k-k^\prime\rightarrow 0$ by the Ward identity.
In the following, these transverse contributions are neglected
and the dressed quark-photon vertex is taken to be
$\Gamma^{\rm BC}_{\mu}(k,k^{\prime})$.  

Calculating the quark-loop integration in \Eqs{LambdaKU} and
(\ref{LambdaKS}), one  obtains the $K$-meson EM form factors and
associated charge radii. Their values are given in
Table~\ref{Tab:BSAmpParam}.

%.........................................................................
%.........................................................................
%
\subsection{Vector meson BS amplitudes}
The model vector-meson BS amplitudes, to be used in Sec.~\ref{Sec:VEMFF}
to calculate the EM form factors of the $\rho^+$, $K^{*+}$ and $K^{*0}$
vector mesons, are similar to the pseudoscalar amplitudes discussed
above.  The Ansatz for the vector-meson BS amplitude $V_\mu(k;p)$ is
\begin{equation}
  V_\mu(k ; p)
  = \T_{\mu \nu}(p) \: \gamma_{\nu} \:
     \frac{ B_{\cal V}(k^2) }{N_{\cal V}} 
,
  \label{VectorBSAmp}     
\end{equation}
where $\T_{\mu \nu}(p) = \delta_{\mu \nu} + p_{\mu} p_{\nu} / M^2$ and 
$M$ is the mass of the vector meson ${\cal V}$.
The BS normalization $N_{\cal V}$ is obtained from the BS normalization
equation, which is a simple generalization of \Eq{fK}.

The difference between the pseudoscalar- and vector-meson BS amplitudes is
provided for by allowing $b^{\cal V}_0$, $b^{\cal V}_1$ and $b^{\cal V}_3$
to differ from the values $b^{\cal P}_0$, $b^{\cal P}_1$ and $b^{\cal
P}_3$.
This simple Ansatz supplies a minimal difference that is sufficient for the
purposes of this present study of the low-$q^2$ behavior of the EM form
factors.   
For large values of $q^2$, one expects that other Dirac
structures, not included in \Eq{VectorBSAmp}, may be important. 
This is indeed what is observed in studies of pseudoscalar mesons, where
the term proportional to $\gamma_5 \gamma \cdot p$ ($p$ is the total
momentum of the pseudoscalar bound state) must be included in the BS
amplitude to reproduce the correct, asymptotic-$q^2$ behavior of the form
factor \cite{Maris2}.  
Higher Dirac moments may also be important for the calculation of some
hadronic decays.  
For example, in quark-model calculations, the particular
choice of quantum numbers for the $q$-$\bar{q}$ interaction which induces
the $\rho \rightarrow \pi \pi$ decay has a considerable impact on the
coupling constant $g_{\rho\pi\pi}$ \cite{Swanson}.  
In particular, this decay is better described within 
quantum mechanics in terms of an interaction that has 
$^{2s+1} L_J$ = $^3 P_0$ quantum numbers, rather than $^3 S_1$.
The latter being equivalent to a one-gluon-exchange interaction.
Such sensitivity to the quantum numbers could show up in the present,
covariant model, as an enhancement of higher Dirac moments.

Using the simple Ansatz given by \Eq{VectorBSAmp},
it is impossible to obtain simultaneous agreement with the experimental
decay widths for 
$\rho \rightarrow \pi\pi$, $\rho \rightarrow e^+ e^-$ and 
$\rho \rightarrow \gamma \pi$ for any reasonable\footnote
{
The parameters are chosen so that the resulting vector BS
amplitudes are monotonic functions of $k^2$, as observed in numerical
studies. 
}
choice of parameters.
On the other hand, one may obtain a satisfactory fit by relaxing the
constraint of reproducing the experimental decay width for $\rho 
\rightarrow \pi\pi$.
This may, in fact, support the notion that this process is particularly
sensitive to higher-order Dirac structures in the BS amplitude which have
been neglected in \Eq{VectorBSAmp}.  
However, a complete study of the importance of higher Dirac moments would
require a careful analysis of the vector-meson BS equation, its dependence
on the 8 possible Dirac structures present in the full vector-meson BS
amplitude, and the extent to which higher Dirac moments affect
vector-meson observables.   
The present investigation of vector-meson EM form factors is a necessary
prerequisite of such a study. 

The magnitude of $b_0^{\rho}$ is closely related to the
$\rho$-meson EM charge radius and should take a value 
$0 \leq b_0^{\rho} \leq b_0^{\pi}$ to ensure the phenomenologically
desirable result:   
$\langle r_{\rho}^2 \rangle \leq \langle r_{\pi}^2 \rangle$.
On this domain, varying the value of $b_0^{\rho}$ 
affects the $\rho$-meson charge radius by less than 10\%  
and the median value of $\langle r_{\rho}^2 \rangle$ occurs when 
$b_0^{\rho} \approx \sfrac{1}{3} b_0^{\pi}$.  
The constraint $b_0^{\cal V} = \sfrac{1}{3} b_0^{\cal P}$,
where ${\cal V}$ = $\rho$ ($K^*$) and ${\cal P}$ = $\pi$ ($K$), 
is employed here, as a means to minimize the number of free parameters in
\Eq{VectorBSAmp}.  
The remaining parameters, $b^{\rho}_1$ and $b^{\rho}_3$ are then
obtained from a $\chi^2$ fit to the decay constants: $f_{\rho}$ and
$g_{\gamma \rho \pi}$, and $b_1^{K^*}$ and $b_3^{K^*}$ are fit to
$g_{\gamma K^* K}$.  The resulting parameters are summarized in
Table~\ref{Tab:QuarkParam} and the calculated decay 
constants and their experimental values are given in 
Table~{\ref{Tab:RhoParam}}.

The formulae necessary to calculate $\rho^+ \rightarrow \gamma \pi^+$ are
described in \Ref{Tandy}, while those necessary to calculate $\rho^0
\rightarrow e^+ e^-$ and $\rho \rightarrow \pi \pi$ are discussed in
\Ref{Pichowsky}.  The decay constant $g_{\gamma K^* K}$ for the $K^*
\rightarrow \gamma K$ decay is obtained from an obvious generalization of
the definition for $g_{\gamma \rho \pi}$ in \Ref{Tandy}, with the
observation that isospin invariance requires:
\begin{equation}
\Gamma_{K^{*+}\rightarrow K^+ \pi^0} = 
\frac{1}{3} \Gamma_{K^{*+}\rightarrow K \pi} \approx 16.6 \pm 0.3 {\rm MeV}
\label{K*+Width}
.
\end{equation}
The experimental value of the coupling constant calculated from
\Eq{K*+Width} is $g_{K^* K\pi} = 6.40 \pm 0.06$.

%
%.........................................................................
%.........................................................................
\section{Electromagnetic form factors of vector mesons}
\label{Sec:VEMFF}
The EM current  of a vector meson is 
\begin{equation}
\langle -p^\prime \lambda^\prime | J_{\mu}(q) | p \lambda \rangle
= \Lambda_{\mu \alpha \beta}(p,p^\prime)
\; \varepsilon_{\alpha}(p,\lambda) 
\;\varepsilon^*_{\beta}(-p^\prime,\lambda^\prime)\:,
\label{Current}
\end{equation}
where $q = - p^{\prime} - p$ is the photon momentum
and   $\varepsilon_{\alpha}(p,\lambda)$ is the polarization vector
      of a vector meson with helicity $\lambda$.
Bose symmetry and charge conjugation require that the proper vertex 
satisfies 
$\Lambda_{\mu \alpha \beta}(p,p^{\prime}) = 
-\Lambda_{\mu \beta \alpha}(p^{\prime},p)$
so that the most general Lorentz covariant form of 
$\Lambda_{\mu \alpha \beta}(p,p^{\prime})$ is
\begin{equation}
    \Lambda_{\mu \alpha \beta}(p,p^{\prime}) =
   - \sum_{j=1}^{3} T^{[j]}_{\mu \alpha \beta}(p,p^{\prime}) F_j(q^2) \:,
   \label{FiDef}
\end{equation}
where
\begin{eqnarray}
  T^{[1]}_{\mu \alpha \beta}(p,p^{\prime}) & = &
    (p_\mu - p^{\prime}_{\mu})\T_{\alpha \gamma}(p) 
\T_{\gamma \beta}(p^{\prime}) \;,\nn \\
  T^{[2]}_{\mu \alpha \beta}(p,p^{\prime}) & = &
     \T_{\mu \alpha}(p) \T_{\beta \delta}(p^{\prime}) p_\delta
    \- \T_{\mu \beta}(p^{\prime}) \T_{\alpha \delta}(p) 
	p^{\prime}_\delta  \:,  
\label{TiDef} \\
  T^{[3]}_{\mu \alpha \beta}(p,p^{\prime}) & = &
     \frac{1}{2 M^2} (p_\mu - p^{\prime}_\mu) 
		\T_{\alpha \gamma}(p) p^{\prime}_\gamma
		 \T_{\beta \delta}(p^{\prime}) p_\delta \;,  \nn
\end{eqnarray}
and $\T_{\mu \nu}(p)$ is defined below \Eq{VectorBSAmp}.
An alternative set of tensors
whose vector-meson Lorentz indices transform irreducibly
under spatial rotations in the Breit frame
may be constructed from linear combinations of these \cite{Gross}. 
The form factors associated with this set of {\it irreducible tensors}
are the electric-monopole, magnetic-dipole and electric-quadrupole
form factors, $G_E$, $G_M$, and $G_Q$, respectively, 
and in our notation, 
they are related to $F_1$, $F_2$, and $F_3$ according to
\begin{eqnarray}
  G_E & = & F_1 + \frac{2}{3} \; \frac{q^2}{4 M^2} G_Q\:, 
	\nn
  \\
  G_M & = & - F_2\:,  \label{FFRel}
  \\
  G_Q & = & F_1 + F_2 + (1 + \frac{q^2}{4 M^2}) F_3\:. 
	\nn
\end{eqnarray}
\nopagebreak
%.............................................................
\subsection{Impulse approximation}

%%............................................................
\begin{figure}[t]
\centering{\ 
\mbox{\ 
 \epsfig{figure=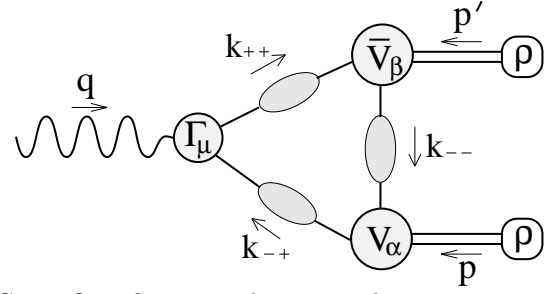,width=7.0cm}}}
\caption{
One of two contributions to the proper $\gamma \rho \rho$ vertex.
}
\vspace{1.0ex}
\label{Fig:GRhoRho}
\end{figure}
%%............................................................
In impulse approximation, the proper $\gamma K^{*+} K^{*-}$ vertex is
given by 
\begin{equation}
\Lambda^{K^{*+}}_{\mu \alpha \beta}(p,p^{\prime}) = 
 \sfrac{2}{3} \Lambda_{\mu \alpha \beta}^{u\bar{s}}(p,p^\prime)
+ \sfrac{1}{3} \Lambda_{\mu \alpha \beta}^{\bar{s}u}(p,p^\prime),
	\label{LambdaDef}
\end{equation}
with
\begin{eqnarray}
  \Lambda_{\mu \alpha \beta}^{u\bar{s}}(p,p^\prime) &=&
  2 N_c {\rm tr}\int \frac{{\rm d}^4k}{(2\pi)^4}
  S_u(k_{++}) i \Gamma_{\mu}(k_{++},k_{-+}) 
\nn \\   & & \times 
  S_u(k_{-+}) \bar{V}_{\beta}(k\-\eta_{+}q;\-p^\prime) 
\nn \\   & & \times 
  S_s(k_{--}) {V}_{\alpha}(k;p) , \label{LambdaU}\\
   \Lambda_{\mu \alpha \beta}^{\bar{s}u}(p,p^\prime) &=&
   2 N_c {\rm tr}\int \frac{{\rm d}^4k}{(2\pi)^4} 
   S_s(k_{++}) i \Gamma_{\mu}(k_{++},k_{-+})
\nn \\   & & \times 
   S_s(k_{-+}) {V}_{\alpha}(k\-\eta_{-}q;-p^\prime) 
\nn \\   & & \times 
S_u(k_{--}) \bar{V}_{\beta}(k;p), \label{LambdaS}
\end{eqnarray}
where 
$k_{ab} = k \+ a \eta_{a}q \+ b \eta_{b} p$ in \Eq{LambdaU},
$k_{ab} = k \+ a {\eta}_{-a}q \+ b {\eta}_{-b} p$ in \Eq{LambdaS},
$\eta_{+} + \eta_- = 1$, 
$\eta_+$ is the fraction of total bound-state momentum carried by the
$u$ quark, 
and 
$V_{\alpha}(k;p)$ is the vector-meson BS amplitude.
The adjoint vector-meson BS amplitude is given by
$\bar{V}_{\alpha}(k;-p) = C^T V_{\alpha}^T(-k;-p) C$, where 
$ C = \gamma_2 \gamma_4$ and $T$ denotes the transpose of Dirac indices.
The proper $\gamma \rho^+ \rho^-$ vertex is given by 
Eqs.~(\ref{LambdaDef})--(\ref{LambdaS})
with the relabeling $s \rightarrow d$
and the proper $\gamma K^{*0} \bar{K}^{*0} $ vertex is given by 
Eqs.~(\ref{LambdaDef})--(\ref{LambdaS})
with the relabeling $u \rightarrow d$  
and the replacement of the coefficient $\sfrac{2}{3}$ by $- \sfrac{1}{3}$ 
in \Eq{LambdaDef}.
The diagram corresponding to \Eqs{LambdaU} and (\ref{LambdaS}) is shown in
\Fig{Fig:GRhoRho}.  

%.........................................................
%............................................................
\begin{table}
\begin{center}
\begin{tabular}{llll}
% 	\hline
 & this study & rel.\ QM \protect\cite{deMelo} & experiment \\
	\hline
$f_\rho$      		& \ 4.37      &&\ 5.05 $\pm$ 0.12 \\
$g_{\gamma\rho\pi}$     & \ 0.62      &&\ 0.57 $\pm$ 0.03 \\
$g_{\gamma K^* K}$      & \ 0.98      &&\ 0.75 $\pm$ 0.03 \\ 
$g_{\rho \pi \pi}$      & \ 8.52      &&\ 6.03 $\pm$ 0.02 \\
$g_{K^{*} K \pi}$       & \ 9.66      &&\ 6.40 $\pm$ 0.04 \\	 \hline
$\langle r_{\rho^+}^2\rangle^{{1}/{2}}$&\ 0.61 fm
	&\ 0.61 $\pm$ 0.014 fm &\ ($\sim$ 0.61 fm) \\
$\langle r_{K^{*+}}^2 \rangle^{{1}/{2}}$&\ 0.54 fm & &         \\
$\langle r_{K^{*0}}^2 \rangle$               & -0.048 fm$^2$&&     \\
$\mu_{\rho^+}$ 
%   ${\scriptstyle[{e_0}/{2M_{\rho}}]}$    
   &\ 2.69 &\ 2.23$\pm$ 0.13 &      \\
$\mu_{K^{*+}}$ 
%   ${\scriptstyle[{e_0}/{2 M_{K^*}}]}$ 
   &\ 2.37   &&      \\
$\mu_{K^{*0}}$ 
%   ${\scriptstyle[{e_0}/{2 M_{K^*}}]}$ 
   &-0.40   &&      \\
$\bar{Q}_{\rho^+}$    &\ 0.055 fm$^{2}$   
       &\ 0.048 $\pm$ 0.012 fm$^2$ &      \\
$\bar{Q}_{K^{*+}}$  &\ 0.029 fm$^{2}$   &&      \\
$\bar{Q}_{K^{*0}}$  & -0.0049 fm$^{2}$ &&      \\
%	\hline
\end{tabular}
\caption
{
Calculated vector meson observables.
The quoted magnetic moments are the dimensionless values $G_M(0)$ 
where $\mu = G_M(0) \frac{e_0}{2 M}$,
quadrupole moments are $\bar{Q} = - G_Q(0) \frac{1}{M^2}$ 
where $M$ is the mass of the vector meson.
The ``experimental'' value of $\langle r^2_{\rho^+} \rangle^{1/2}$ 
is estimated from diffractive electroproduction data 
\protect\cite{Pichowsky2} and results from relativistic quantum mechanics
are provided \protect\cite{deMelo}.
}
\label{Tab:RhoParam}
\end{center}
\end{table}
%............................................................
%.........................................................
%
%%............................................................
\begin{figure}[t]
\centering{\ 
\mbox{\ 
\epsfig{figure=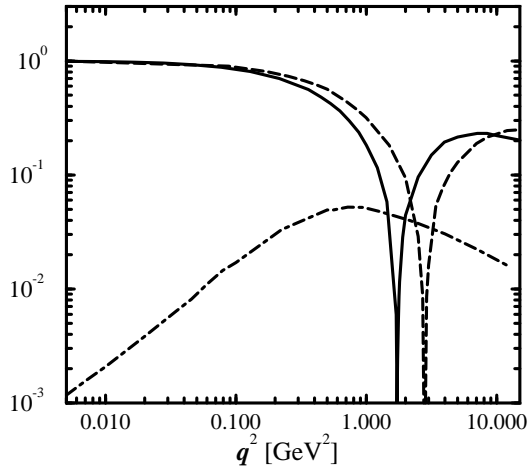,width=8.0cm}}}
\caption{
The electric-monopole form factor $|G_E(q^2)|$ for
$\rho^+$ (solid), $K^{*+}$ (dashed) and $K^{*0}$ (dot-dashed) mesons.  
The sign of $G_E(0)$ is positive for each of the mesons considered.
% $\rho^+$ and $K^{*+}$ mesons and negative for $K^{*0}$. 
}
\label{Fig:Ge}
\end{figure}
%%............................................................
%
%.....................................................................
\nopagebreak
\subsection{Numerical results}
\begin{figure}[t]
\centering{\ 
\mbox{\ 
\epsfig{figure=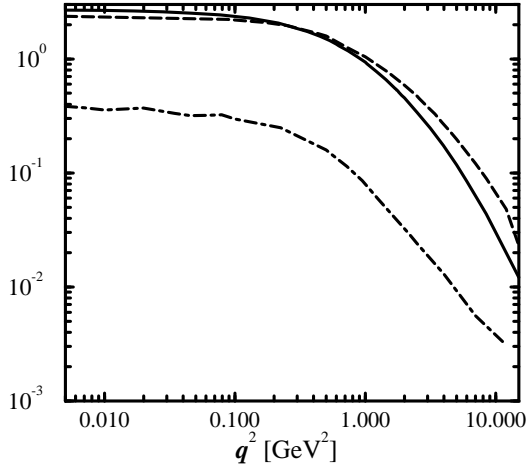,width=8.0cm}}}
\caption{
The magnetic-dipole form factor $|G_M(q^2)|$ for
$\rho^+$ (solid), $K^{*+}$ (dashed) and $K^{*0}$ (dot-dashed) mesons.  
The sign of $G_M(0)$ is positive for $\rho^+$ and $K^{*+}$ and
negative for $K^{*0}$.
}
\label{Fig:Gm}
\end{figure}

\begin{figure}[t]
\centering{\ 
\mbox{\ 
\epsfig{figure=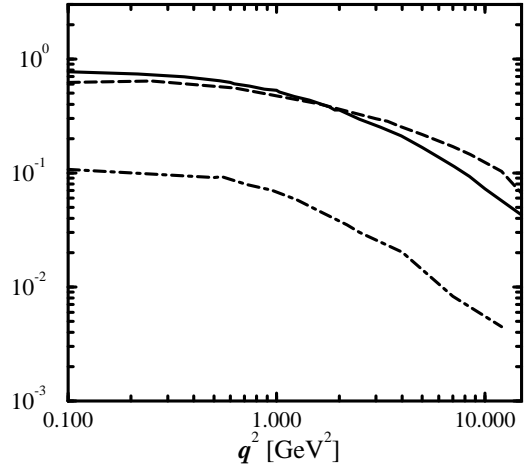,width=8.0cm}}}
\caption{
The electric-quadrupole form factor $|G_Q(q^2)|$ for
$\rho^+$ (solid), $K^{*+}$ (dashed) and $K^{*0}$ (dot-dashed) mesons.  
The sign of $G_Q(0)$ is negative for $\rho^+$ and $K^{*+}$ and
positive for $K^{*0}$.
}
\label{Fig:Gq}
\end{figure}

Having determined the parameters in the vector-meson BS amplitude,
one may calculate the EM form factors directly
from Eqs.~(\ref{LambdaU}) and (\ref{LambdaS}).  
The resulting form factors are shown in \Figs{Fig:Ge} -- \ref{Fig:Gq}.

The nodal structures of the form factors are typical of spin-1 systems.
$F_1(q^2)$ has a node at moderate $q^2$, $F_3(q^2)$ has
a node at a higher $q^2$ and $F_2(q^2)$ has no nodes.
From the relations in Eqs.~(\ref{FFRel}), one finds that $G_E(q^2)$ has
then two nodes, $G_Q(q^2)$ has one node and $G_M(q^2)$ has none.  
The node in $G_Q(q^2)$ and second node in $G_E(q^2)$ occur above $q^2$ =
10~GeV$^2$ hence they are not depicted in \Figs{Fig:Ge} -- \ref{Fig:Gq}.
The location of the first node in $G_E(q^2)$ gives a measure of the size
of the vector meson.  Its appearance at a smaller $q^2$ for the
$\rho^+$ meson relative to that of the $K^{*+}$ is indicative of the
larger size of the $\rho^+$.

%%............................................................
%%............................................................

The canonical measure of the size of a bound state is the charge radius,
defined in \Eq{Def:Radius}.
The charge radius of the $\rho^+$ meson has also been calculated within 
relativistic quantum mechanical models \cite{Keister,deMelo}.   
The mean charge radius obtained from and the deviation between these
models is $\langle r_{\rho^+}^2 \rangle^{1/2}$ = 0.61 $\pm$ 0.014~fm. 
The result of the present model is
$\langle r_{\rho^+}^2 \rangle^{1/2}=$ 0.61~fm. 

It is important to note that the charge radii that result from our
calculations of Eqs.~(\ref{LambdaU}) and (\ref{LambdaS}) represent only
the contributions due to the quark-antiquark substructure of the vector
meson.  
Additional contributions arising from meson loops may be present but have
been neglected in this study and in Refs.~\cite{Keister,deMelo}. 
In the case of the pion form factor, the effect of including $\pi$-meson
loops is minimal, tending to increase the pion charge radius by less than
15\% \cite{Bender}.
However, meson loops have a qualitatively different effect on the
vector-meson form factors.
The light vector mesons are massive enough to decay into two
pseudoscalars.  
Therefore, two-pseudoscalar meson loops in vector-meson self-energies give
rise to nonanalyticities associated with this decay. 
Such nonanalyticities may have significant impact on certain observables,
particularly the charge radii of {\em neutral} particles.
An example of this is the semileptonic decays 
$K \rightarrow \pi \ell \nu_\ell$ and $\pi \rightarrow \pi \ell \nu_\ell$.
It was shown that subthreshold $K\pi$ loops contribute significantly
to the quark-antiquark-$W$-boson vertex for {\em timelike} $W$-boson
momenta \cite{Kalinovsky}. 
The most significant effect on an observable being an increase in obtained
radius for the $\pi$-$K$ transition form factor by up to 50\%.
Nonetheless, since such contributions fall off rapidly with increasing
spacelike momenta, they are expected to have a smaller effect on the
{\em spacelike} EM form factors considered here. 
The role and importance of pion loops in determining the $\rho$-meson EM
form factors for timelike and spacelike photon momenta requires further 
investigation.

Although a direct measurement of the $\rho$-meson charge radius is not
possible at present, there is an empirical relation between
the $t$ dependence of diffractive $\rho$-meson electroproduction and a 
``diffractive radius'' of the vector meson.   
The ``diffractive radius'' that is inferred from electroproduction
data is $r_\rho \approx$ 0.61~fm \cite{Pichowsky2}.  
However, the agreement between the ``diffractive radius''
and the charge radii obtained in Refs.~\cite{Keister,deMelo} and herein should
be viewed with caution since the relationship between these quantities is
not understood. 

It is apparent from \Figs{Fig:Ge} -- \ref{Fig:Gq} that the neutral $K^*$
meson form factor is not identically zero for all $q^2$.  
In general, the EM form factors of charge-conjugate non-identical,
neutral mesons (such as the $K^{*0}$ and $\bar{K}^{*0}$ mesons) will not
be identically zero for all $q^2$. 
This results from the breakdown of SU(3)-flavor symmetry caused by the
mass difference between the $u$ and $s$ quarks.  
The negative value of $\langle r_{K^{*0}}^2 \rangle$ is indicative of the
larger mass of the $s$ quark.
Of course, the EM form factors for the neutral-$\rho$ meson are absent
from the present study because the assumption of exact isospin invariance
at the level of the quark dynamics (i.e., $S_u(k) = S_d(k)$)
entails that they are identically zero for all $q^2$.

Of course, the value of $G_E(q^2)$ at $q^2 = 0$ is proportional to the
EM charge of the vector meson. 
In the numerical calculation of a neutral meson form factor, 
a result of $G_E(0) \not = 0$ would be indicative of a violation of charge
conservation in \Eqs{LambdaU} and (\ref{LambdaS}).
In the present study, charge conservation (i.e., $G^{K^*0}_E(0) = 0$)
is ensured in \Eqs{LambdaU} and (\ref{LambdaS}), with the assumption that
the $u$ quark carries a fraction $\eta_+ = 0.45$ of the total bound-state 
momentum $p$ while the $\bar{s}$ quark carries the fraction $1 - \eta_+$.
Had the assumption of equal partitioning $(\eta_+ = 1/2)$ been made, a 
small violation to charge conservation would have ensued. 
This violation would have been $G_E(0) = 0.0032$.

It was shown in \Ref{Maris}, that charge conservation is maintained within
the impulse approximation through 
the use of BS amplitudes, satisfying the Bethe-Salpeter equation, in which
the full dependence on the invariant $k\cdot p$ is maintained.  
(Here, $k$ is the relative quark-antiquark momentum and $p$ is
the total bound state momentum.)
In this study, we have employed simplified model BS amplitudes, having
only the dominant Lorentz structure ${\bf T}_{\mu \nu}(p) \gamma_{\nu}$ 
and arguments that are $O(4)$ symmetric; i.e., independent of $k\cdot p$.
Therefore, charge conservation was ensured through the choice of $\eta_{+}
=$ 0.45 in \Eqs{LambdaU} and (\ref{LambdaS}).

%%............................................................
%%............................................................

Overall, the agreement between the $\rho$-meson
form factors obtained from our model and those obtained
from relativistic quantum mechanical models is quite good for small $q^2$.
Below $q^2 = 1\;{\rm GeV}^2$, both approaches are expected to be reliable 
and the agreement achieved may be indicative of this.  
For example, the mean charge radius and quadrupole moment obtained in
\Ref{deMelo} is 
$\langle r^{2}_{\rho} \rangle^{1/2} = 0.61 \pm 0.014$ fm and
$\bar{Q} =$ 0.048 $\pm$  0.012~fm$^2$ 
while those obtained in the present study are
$\langle r^{2}_{\rho} \rangle^{1/2} = 0.61$ fm and 
$\bar{Q} =$  0.055~fm$^2$, respectively.
However, the discrepancy between the magnetic-dipole moment 
obtained in \Ref{deMelo} $\mu =$ 2.23 $\pm$ 0.13 $[{e_0}/{2 M_{\rho}}]$,
and that obtained in the present study 
$\mu =$ 2.69 $[e_0/2 M_{\rho}]$ is significant. 
An experimental determination of vector-meson magnetic moments
provides a tool with which to probe the various model frameworks.

Furthermore, in the present study, a close relation between the charge
radius and magnetic moment of a vector meson and the infrared behavior of
its BS amplitude is observed.  
A measurement of either of these observables would provide information
about the infrared behavior of the vector-meson BS amplitudes. 
Such information is important, as it provides crucial constraints on
future numerical studies of vector-meson bound states in QCD.

An important difference between quantum mechanical models and
Dyson-Schwinger-based models is that quantum mechanical models typically
employ a constituent quark with a mass $M_{\rho}/2$, while in our  
approach, quarks are confined and have {\em no} mass pole at all.  
The mass-scales in the quark propagator $S_f(k)$ of \Eq{QuarkProp} evolve
dynamically with the quark momentum, being constituent-like near $k^2 = 0$
and becoming current-like as $k^2 \rightarrow \infty$.
Therefore, at low-$q^2$, one expects the two approaches to produce similar
results, and (with the exception of the magnetic moment) the results
obtained herein are in complete agreement with those of \Ref{deMelo}.

However, above $q^2 = 1$ GeV$^2$ the results of relativistic quantum mechanics
begin to differ from those calculated herein.
For example, the location of the first node in $G_E(q^2)$, as obtained in
Refs.~\cite{Keister,deMelo}, occurs at $q^2 \approx $ 3~GeV$^2$, while it
is found from our calculation to be near 1.73~GeV$^2$.
This discrepancy results from two differences
between quantum mechanical models and our Dyson-Schwinger based model.
The first is the use of {\em dynamical} mass-scales in the
quark propagator $S_f(k)$ from \Eq{QuarkProp}, as discussed above.  
The transition in the quark propagator $S_f(k)$ from a constituent-like
quark to a current-like quark becomes important in the calculation of EM
form factors for photon momenta greater than $q^2 \approx$ 1~GeV$^2$.   
This feature is common to processes involving photo-meson transitions and
has been observed in studies of diffractive electroproduction
of vector mesons \cite{Pichowsky} and the $\gamma^* \pi \rightarrow \gamma$
transition form factor \cite{Frank:ggp}. 
The second difference is the sensitity of the vector-meson form factors to
the lack of rotational covariance in relativistic quantum mechanical
models.  The uncertainties in the vector-meson form factors associated
with this sensitivity become significant above $q^2 = $ 1~GeV$^2$
\cite{Keister}.   
In contrast to this, the framework employed herein is fully covariant.
%

%.....................................................................
\section{Summary}
The electromagnetic form factors, associated moments and charge radii
have been calculated for the $\rho^+$, $K^{*+}$, and $K^{*0}$ mesons
within a Dyson-Schwinger based model using confined-quark propagators
developed in phenomenological studies of $\pi$- and $K$-meson observables
and vector-meson BS amplitudes that are closely related to the
pseudoscalar amplitudes obtained from the same studies.

The calculated $\rho$-meson charge radius and electric-quadrupole moment
are in excellent agreement with the values obtained from light-front
quantum mechanical models, while the magnetic moment obtained herein is
found to be 20\% larger than the mean value obtained from quantum
mechanical models.

The experimental difficulties that must be overcome to directly measure
the vector-meson EM form factors will continue to limit our knowledge of
these important observables in the future.  
However, indirect measurements of their static limits are possible and
provide important information about quark dynamics and vector-meson bound
states. 
In particular, the quadrupole moment provides an excellent probe of the
mass scales in the quark propagator, while the charge radius principally
probes the details of the vector-meson BS amplitude.
Information about infrared observables provides constraints on and
guidance for the construction of realistic models of the strong interaction
as well as a test of our understanding of nonperturbative QCD.

%.....................................................................
\section*{Acknowledgments}
The authors acknowledge useful conversations with 
S.C.~Capstick, J.~Piekarewicz, and C.D.~Roberts.
This work is supported by the U.S. Department of Energy
under Contracts DE-AC05-84ER40150 and DE-FG05-92ER40750 and 
the Florida State University Supercomputer Computations Research Institute
which is partially funded by the Department of Energy through 
Contract DE-FC05-85ER25000.
%.....................................................................
%.....................................................................


\begin{references}
\bibitem{Pichowsky2} 
	The ``diffractive radius'' of the $\rho$ meson is given by:
	$r_{\rho}^2 = 4 b_{\rho N}(W) - 2 b_{N N}(W) 
 	- 4 \alpha_1 \ln(\alpha_1 W^2)$,
	where $W \geq$ 10~GeV is the CM energy, $\alpha_1 = $ 0.25~GeV$^2$
	is the slope of the Pomeron trajectory, and
	$b_{\rho N}$ and $b_{NN}$ are the slopes of the differential
	cross sections at $t = 0$ for $\rho$-meson electroproduction on a
	nucleon and nucleon-nucleon scattering, respectively. 
	This result relies on the presumed analogy between these exclusive,
	diffractive processes and absorptive scattering in quantum
	mechanics.
\end{references}
\end{document}